\documentclass[twocolumn,3p]{elsarticle}

\usepackage{graphicx}
\usepackage{dcolumn}
\usepackage{bm}
\usepackage{color}
\usepackage{pdfpages}

\begin{document}

\title{An above-barrier narrow resonance in $^{15}$F}

\author{F. de~Grancey$^1$, A. Mercenne$^{1}$, F. de~Oliveira~Santos$^{1}$, T. Davinson$^2$, O. Sorlin$^1$, J.C. Ang\'{e}lique$^3$, M. Assi\'{e}$^{1,4}$, E. Berthoumieux$^5$, R. Borcea$^8$, A. Buta$^8$, I. Celikovic$^{7}$, V. Chudoba$^9$, J.M. Daugas$^6$, G. Dumitru$^8$, M. Fadil$^1$, S. Gr\'{e}vy$^{1,10}$, J. Kiener$^11$, A. Lefebvre-Schuhl$^{11}$, N. Michel$^{1}$, J. Mrazek$^{9}$, F. Negoita$^8$, J. Oko{\l}owicz$^{12}$, D. Pantelica$^8$, M.G. Pellegriti$^1$, L. Perrot$^{1,4}$, M. P{\l}oszajczak$^1$, G. Randisi$^1$,  I. Ray$^1$, O. Roig$^6$, F. Rotaru$^8$, M.G. Saint~Laurent$^1$, N. Smirnova$^{10}$, M. Stanoiu$^8$, I. Stefan$^{1,4}$, C. Stodel$^1$, K. Subotic$^{7}$, V. Tatischeff$^{11}$, J.C. Thomas$^1$, P. Uji\'{c}$^{7}$, R. Wolski$^{12}$}

\address{
$^1$ GANIL, CEA/DRF-CNRS/IN2P3, Bvd Henri Becquerel, 14076 Caen, France\\
$^2$ Department of Physics and Astronomy University of Edinburgh, Edinburgh EH9 3JZ, United Kingdom \\
$^3$ LPC Caen, ENSICAEN Universit\'e de Caen, CNRS/IN2P3 Caen, France  \\
$^4$ IPN Orsay, France \\
$^5$ CEA Saclay Irfu/SPhN F-91191 Gif-sur-Yvette, CEA,
$^6$ DAM, DIF, 91297 Arpajon cedex, France\\
$^{7}$ Vin\v{c}a Institute of Nuclear Sciences, University of Belgrade Belgrade, Serbia \\
$^8$ Horia Hulubei National Institute of Physics and Nuclear Engineering, P.O. Box MG6 Bucharest-Margurele, Romania \\
$^9$ Nuclear Physics Institute ASCR, CZ-25068 Rez, Czech Republic \\
$^{10}$ UMR 5797, CNRS/IN2P3, Universit\'e de Bordeaux, Chemin du Solarium, 33175 Gradignan Cedex, France \\
$^{11}$ CSNSM, CNRS/IN2P3/Universit\'e Paris-Sud, B\^at.~104 91405 Orsay Campus, France \\
$^{12}$ Institute of Nuclear Physics, PAS, Radzikowskiego 152, PL-31342 Krak\'ow, Poland \\

}


\begin{abstract}
Intense and purified radioactive beam of post-accelerated $^{14}$O was used to study the low-lying states in the unbound $^{15}$F nucleus. Exploiting resonant elastic scattering in inverse kinematics with a thick target, the second excited state, a resonance at E$_R$=4.757(6)(10)~MeV with a width of $\Gamma$=36(5)(14)~keV was measured for the first time with high precision. The structure of this narrow above-barrier state in a nucleus located two neutrons beyond the proton drip line was investigated using the Gamow Shell Model in the coupled channel representation with a $^{12}$C core and three valence protons. It is found that it is an almost pure wave function of two quasi-bound protons in the $2s_{1/2}$ shell.
\end{abstract}



\maketitle

%
%
\section{Introduction}
The boundaries for nuclear stability against particle emission are called "drip lines". Beyond the drip lines the particle emission time $\tau$ is usually shorter than ~10$^{-21}$ seconds and unbound nuclei are observed as broad resonances. The proton drip line plays an important role in nuclear astrophysics, e.g. in the Main Sequence and Red Giant stars, during the Big Bang Nucleosynthesis ($^{2}$He, $^4$Li, $^5$Li, $^8$Be are unbound) and in the $rp$-process \cite{Fisker08} during type~I X-ray bursts where rapid proton captures reactions reaching the proton drip line (from $^{15}$F up to $^{101}$Sb) should wait for $\beta$ decays before proceeding further.
\par
However, formation of narrow resonances at high excitation energies and beyond drip mines is not an unexpected phenomenon. If the spacing between the resonances becomes smaller than their widths, the salient reordering processes under the influence of the environment of continuum states takes place. This phenomenon which is related to the avoided crossing in the complex energy plane has been refereed to as resonance trapping \cite{klein,zelevinsky03} and proved experimentally on an open microwave billiard \cite{persson}. In atomic nuclei, direct studies of resonance trapping are not feasible because one cannot trace widths of states as a function of the coupling strength to continuum. However, much information have been accumulated indirectly which contradict the naive expectation based on the random matrix theory that all nuclear levels will be broadened with increasing coupling strength to the continuum \cite{Oko1}.
\par
Narrow resonances are also known at low excitation energies in the domain of isolated resonances.
 Particularly interesting are those which appear due to either the avoided crossing of resonances, or the near-threshold collectivization in the ensemble of shell model (SM) states which yields the collective state which carries many features of the nearby particle emission threshold. A well known example is the unbound nucleus $^8$Be which decays in the ground state by 2$\alpha$ particles. The proximity of 2$\alpha$-decay threshold enhances the 2$\alpha$ correlations in the ground state $0_1^+$ wave function and imprints the nature of $2_1^+$ and $4_1^+$ broad resonances \cite{Okolowicz12,Okolowicz13}.
 Interestingly, at higher excitation energies in the vicinity of one-proton (1p) and one-neutron (1n) decay thresholds, one finds a group of narrow resonances $(2_2^+ - 2_3^+)$,  $(1_1^+ - 1_2^+)$, and $(3_1^+ - 3_2^+)$ which carry the imprint of nearby decay channels $[{^7}$Li-p] and $[{^7}$Be-n] and hence, the 2$\alpha$ component in these resonances is suppressed. As a result, one observes a strong reduction of the decay width for these states as compared to the width of $2_1^+$ and $4_1^+$ resonances. The coupling to these channels is also at the origin of width attraction
 $(\Gamma_{i_1}\simeq\Gamma_{i_2})$ in $(2_2^+ - 2_3^+)$ and $(3_1^+ - 3_2^+)$ doublets of resonances, and the width repulsion $(\Gamma_{i_1}<<\Gamma_{i_2})$ in $(1_1^+ - 1_2^+)$ resonances \cite{Brentano}.  Width attraction and width repulsion are two generic features of the avoided crossing of resonances in open quantum systems \cite{Oko1}.
\par
It is interesting to compare these results for $^8$Be with others open quantum systems to separate generic aspects of the continuum coupling from specific aspects which are related to the nuclear interaction, such as the ordering and the excitation energy of the particle emission thresholds. In this work, we study the spectroscopy of $^{15}$F, the isotope of fluorine located two neutrons beyond the proton drip line. The reordering processes induced by the continuum couplings in this nucleus are governed  by different sets of channels than in $^8$Be. The present work reports on the observation of a narrow state in the vicinity of the 2p emission threshold using the resonant elastic scattering reaction $^{14}$O(p,p)$^{14}$O measured in inverse kinematics.  The structure of this state was also investigated using the Gamow Shell Model in the coupled channel representation with a $^{12}$C core plus three valence protons.

\section{Status of $^{15}$F}

Properties of the ground state (J$^{\pi}$~=~$1/2_1^+$) and the first excited state (J$^{\pi}$~=~$5/2_1^+$) of $^{15}$F were measured several times \cite{Kekelis78,Benenson78,Peters03,Lepine03,Lepine04,Goldberg04,Guo05}, see Ref.\cite{Fortune06} for a compilation of the results. The ground state is unbound by $\approx$~1.3~MeV and is observed as a broad resonance with $\Gamma \approx$~0.5-1.3~MeV \cite{Fortune06,Mukhamedzhanov10,Baye05}. The first excited state is unbound by $\approx$~2.8~MeV and is observed as a narrower resonance of $\Gamma \approx$~300~keV. Both states are well described as single-particle configurations with dimensionless reduced width, sometimes called spectroscopic factor, $\theta^2 > $~0.5  \cite{Fortune06,Fortune05,Grigorenko15}. The structure of the ground (first excited) state is interpreted as mainly a proton orbiting with $\ell$~=~0 ($\ell$~=~2) around a $^{14}$O$_{gs}$ core \cite{Fortune05}.

\par
Candidate for the second excited state can be looked at in the mirror nucleus. The second excited state in the mirror nucleus $^{15}$C is known at the energy of 3103~keV, with J$^{\pi}=1/2_1^{-}$ and a width $\Gamma=29(3)$~keV \cite{Fortune11}. This state was populated strongly in two-neutron transfer reactions with a $^{13}$C target \cite{Truong83,Cappuzzello12}, indicating a structure of mainly two sd-shell neutrons coupled to a $^{13}$C core. Canton \emph{et al.} \cite{Canton06} used the multichannel algebraic scattering theory with Pauli-hindered method in order to predict the properties of the low-lying states in $^{15}$F.  A very narrow width $\Gamma$=5~keV was predicted for the second excited state, see Table \ref{table1}. 
Fortune and Sherr \cite{Fortune07} used a potential model to determine the single-particle widths which they scaled down to reproduce the measured widths in $^{15}$C. The extracted $\theta^2$ were used to calculate widths in the mirror nucleus $^{15}$F. These calculations confirmed that narrow resonances are to be expected in $^{14}$O+p, but they obtained a width 10~times larger than the one of Ref. \cite{Canton06} for the second excited state. Refined values were later published by Fortune \cite{Fortune11}, see Table \ref{table1}. Canton \emph{et al.} \cite{Canton07} objected that $\theta^2$ do not necessary scale with the single-particle widths, especially when the $\theta^2$ is small \cite{Oliveira97}.

\par
A first indication for the observation of the second excited state in $^{15}$F was obtained at GANIL by Lepine-Szily \emph{et al.} \cite{Lepine04} through the measurement of the transfer reaction $^{16}$O($^{14}$N,$^{15}$C(0.740 MeV)$^*$)$^{15}$F. A narrow peak of $\approx$10~counts with a width of only 150(100) keV was observed. This state was also observed through the angular correlations of decay products in the fragmentation of $^{17}$Ne \cite{Mukha09}, with slightly more statistics ($\approx$20~counts) but a worse resolution. Results of these measurements are summarized in Table \ref{table1}.  The narrow width is particularly surprising since this state is located 3.5~MeV above the Coulomb plus centrifugal barrier of the system $^{14}$O+p. The spin of this second excited state has not been assigned.

\begin{table}
\caption{\label{table1} Resonance energy, width and spin measured and theoretical predictions for the second excited state of $^{15}$F.}
\vskip 0.3truecm
\begin{tabular}{|c|c|ccc|}
\hline
\multicolumn{1}{|c|}{} & \multicolumn{1}{|c|}{Ref.} &  \multicolumn{3}{c|}{Second excited state}\\
\multicolumn{1}{|c|}{} & \multicolumn{1}{|c|}{} &E$_R$(MeV)&$\Gamma$(keV)&J$^\pi$ \\
\hline
Measured&\cite{Lepine04}&4.800(100)&150(100)&-\\
&\cite{Mukha09}&4.900(200)&200(200)&-\\
&Present&4.757(16)&36(19)&$\frac{1}{2}^-$\\
\hline
Predicted&\cite{Canton06}&5.49&5&$\frac{1}{2}^-$\\
&\cite{Fortune07}&4.63&55&$\frac{1}{2}^-$\\
&\cite{Fortune11}&4.63&38&$\frac{1}{2}^-$\\
\hline
\end{tabular}

\end{table}

\section{Experiment}
\label{Experiment}
The $^{15}$F nucleus was studied using the resonant elastic scattering technique. The excitation function of the elastic scattering reaction $^{14}$O(p,p)$^{14}$O was obtained in inverse kinematics using a thick target. The excitation function at low energy is dominated by Coulomb scattering, but it also shows peaks and interferences that correspond to the presence of resonances in the compound nucleus. The properties of these resonances, i.e. resonance energy, width, and spin, can be extracted from the analysis of the shape of the peaks using the R-Matrix formalism \cite{Descouvemont10,azure,Baye05}. More details on the procedure can be found in Ref. \cite{Axelsson96,Oliveira05,Assie12,Stefan14}. The experiment was performed at the GANIL SPIRAL1 facility. Two beams were used: the radioactive beam of $^{14}$O for the study of $^{15}$F, and a stable beam $^{14}$N for calibrations. Radioactive $^{14}$O$^{3+}$ ions were produced through the fragmentation of a 95~MeV/u $^{20}$Ne primary beam impinging on a thick carbon production target. The ions were post-accelerated with the CIME cyclotron up to the energy of 5.95(1)~MeV/u with an energy spread $<$0.2\%. The isobaric contamination of the beam was reduced down to 0.0(1)~\% using a 0.9~$\mu$m thick stripper aluminium foil located at the entrance of the LISE zero degree achromatic spectrometer. The $^{14}$O$^{8+}$ ions were selected using LISE and transported to the experimental setup located in the D4 experimental area. An average beam intensity of 1.88(1)x10$^5$~pps was achieved. This value was obtained by regularly measuring the beam intensity with a silicon detector in conjunction with a calibrated beam intensity reduction system, as well as by counting the 2.312~MeV $\gamma$-ray emitted in the $\beta$-decay of $^{14}$O using a high-purity germanium detector.

\par
The beam was sent to a thick target where it was stopped. The target was made of three (four in the case of $^{14}$N) polypropylene (CH$_{2}$)$_{n}$ foils, 50~$\mu$m thick each. The foils were fixed side by side with the last one put on a 250~rpm rotating system called FULIS \cite{Stodel04}. This system was used to reduce the background arising from the $\beta$-delayed proton emission of $^{14}$O (t$_{1/2}$=70.6~s). Counting rate was reduced from 85~Hz with the stopped target to $\approx$1~Hz with the rotating target. The scattered protons were detected downstream in a $\Delta$E(500~$\mu$m)-E(6~mm cooled SiLi) telescope of silicon detectors that covered an angular acceptance of $\pm$2.2(2)$^\circ$. Identification of the protons was made using contours on $\Delta$E-E and time-of-flight parameters.

\par
The elastic scattering reaction $^{14}$N(p,p)$^{14}$N was measured under the same experimental conditions to calibrate the detectors in energy. Energy calibration and resolution were obtained by populating known resonances in the compound nucleus $^{15}$O  \cite{Olness58, West69} using the same procedure as discussed in Ref. \cite{Guo05}. An experimental energy resolution, $\sigma_{c.m.}$~=~7(2)~keV, was measured from the width of the observed peaks and using an alpha source. No change was measured as a function of the proton energy. The major contributions to this resolution were from the $\Delta$E-E detectors (4.2~keV and 3.0~keV) and the beam and proton straggling in the target (4.7~keV).

\par
The polypropylene target also contains carbon atoms, which induced a background through reactions with the beam. This carbon-induced proton background was not measured in the present experiment. Instead the results published in Ref. \cite{Guo05} obtained in very similar experimental conditions were used. This background was normalized according to carbon content and beam intensity, and subtracted to the measured proton spectrum. The proton background was featureless, almost flat with a weak maximum of 0.09~barn/sr at 1.8~MeV~(c.m.) \cite{Guo05,Degrancey09}.

\section{Results}
\label{Results}

 \begin{figure*}
 \center
\resizebox{0.8\textwidth}{!}{%
\includegraphics{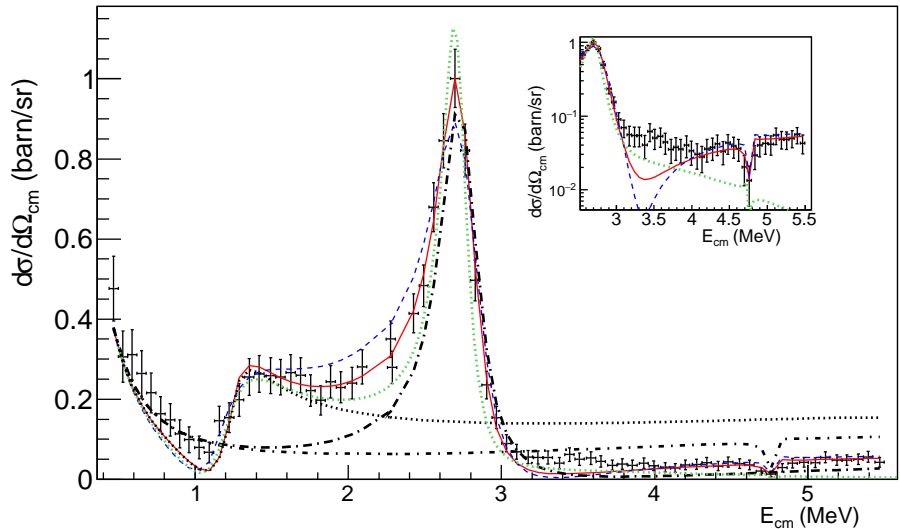}%
}
\caption{(Color online) The excitation function of the reaction $^{14}$O(p,p)$^{14}$O measured at 180$^{\circ}$ in the center of mass system. R-Matrix calculations corresponding to the ground state alone (dotted line), first excited state (thick-dot-dash line), second excited state (thin-dot-dash line) are shown. Inset: A structure clearly visible at an energy close to 4.8~MeV is assigned to the $1/2_1^-$ second excited state of $^{15}$F (see text for details). Data are compared to the best R-Matrix fit (red line) using the properties given in Table \ref{table1}. The R-Matrix calculation made with $\Gamma$~=~737~keV for the ground state is also shown (blue dashed line) for comparison. Here, the error bars correspond to the statistical uncertainties. The calculation using the GCM-CC approach is also shown (green dotted line).}
\label{14Ocm}
\end{figure*}

The measured excitation function of the $^{14}$O(p,p)$^{14}$O reaction, performed at 180$^{\circ}$ (c.m.), is shown in Fig.~\ref{14Ocm}. It is very similar to those obtained in Ref. \cite{Peters03,Goldberg04,Guo05}, but with a much higher statistics, improved energy resolution and covering a larger energy range. An analysis of the excitation function using the R-Matrix method was performed with the code AZURE2 \cite{azure}. A nominal value of the radius parameter $a=5.1$~fm was used.
\par
 A deep minimum is observed at $\approx$~1~MeV corresponding to the well known J$^{\pi}=1/2_1^+$  ground state resonance of $^{15}$F. It is fitted at an energy E$_{R}$~=~1270(10)(10)~keV with $\Gamma$~=~376(70)($_0^{+200}$)~keV, where the quoted uncertainties correspond to statistical and systematic uncertainties respectively. The resonance energy is in agreement with last published value E$_{R}$~=~1230(50)~keV \cite{Guo05}. The measured width is lower than the values obtained in previous studies by at least 30\%, but is still within the error bars. This lower value is also supported by some theoretical considerations \cite{Timofeyuk06,Mccleskey05}. In Fig. \ref{14Ocm}, the best fit is shown with the continuous red line which leads to $\Gamma$~=~376~ keV. For comparison, the dashed-blue line shows the calculation made for the average value of the previous results, i.e. $\Gamma$~=~737~keV. Neither of the two calculations reproduce the data at E$_{c.m.}$~$\approx$~3.4~MeV. The largest differences between the two calculations are found at E$_{c.m.}$~$\approx$~2.2~MeV, where the absolute cross section is most sensitive to the carbon-induced background. Since the carbon-induced background was not measured, a systematic error of $\pm$~25~mbarn/sr on its cross section was adopted for the full range of energy, which resulted in a systematic error for the g.s. width of ($_0^{+200}$)~keV.

\par
  The peak observed at the resonance energy E$_{R}$~=~2763(9)(10)~keV with $\Gamma$~=~305(9)(10)~keV corresponds to the J$^{\pi}=5/2_1^+$ first excited state. It is in good agreement with the previous measurements. The controversy about the width of this state \cite{Fortune06}, resulting in $\theta^2$ exceeding unity, is due to the very small single particle width $\Gamma_{s.p.}$=250~keV calculated in Ref. \cite{Benenson78}. A more meaningful value of $\theta^2=0.42$ is obtained using $\Gamma_{s.p.}=726$~keV calculated with a conventional formula \cite{Rolfs88}. It is possible to calculate a weighted average of all values measured so far using the statistical procedure of the Particle Data Group \cite{Olive14} assuming a Gaussian distribution of the different and independent measurements  \cite{Severijns08}. We obtain the recommended values E$_{R}$~=~2794(16)~keV and $\Gamma$~=~301(16)~keV.

\par
In addition, for the first time in a resonant elastic scattering experiment, the second excited state is clearly observed (insert of Fig.~\ref{14Ocm}) as a narrow dip at a resonance energy of $\approx$~4.8~MeV. In the corresponding mirror nucleus, the second excited state has spin J$^{\pi}=1/2_1^-$. The resonance has the shape predicted by Canton \emph{et al.}  \cite{Canton06}, which is due to destructive interferences between the J$^{\pi}=1/2_1^-$ resonance and Coulomb scattering. No other spin assignment can better reproduce the shape of the structure. The other solutions also induce a peak arising from constructive interferences that is not observed. It is the first time the spin of this state is assigned. The R-Matrix analysis of the excitation function was performed taking into account the experimental resolution. The resonance is measured to be E$_R$~=~4.757(6)(10)~MeV with $\Gamma$~=~36(5)(14)~keV. The measured properties are in good agreement with the previous experimental results, see Table \ref{table1}. The present work shows a significant improvement in the resolution, at least by a factor 5. All resonances predicted at higher energies \cite{Canton06,Fortune07} were taken into account in the R-Matrix fit of the excitation function, but their properties were kept fixed.  Negative parity states are taken from Ref. \cite{Fortune07} and the positive parity states from Ref. \cite{Canton06}. A 500~keV uncertainty in the energy of these higher lying resonances was taken into account in the systematic uncertainties.

\par
The observation of this narrow resonance in $^{15}$F is surprising since this resonance is located well above the Coulomb plus centrifugal barrier ($B_{C}+B_{\ell}~\approx$~3.3~MeV) for the proton emission, there is no barrier to retain the proton inside the nucleus. The single-particle width of this state is 1.6-3.0~MeV \cite{Fortune07} (depending on the model parameters), compared to the measured value of 36(19)~keV. This implies that the measured lifetime is more than 40~times longer. As discussed before, the second excited state is known in the mirror nucleus $^{15}$C at an energy of 3103~keV. It is unbound with respect to one neutron emission, and has a width of $29(3)$~keV \cite{Fortune11}. Here too, the resonance is located above the $\ell=1$ centrifugal barrier ($B_{\ell}~\approx$~1.2~MeV), nevertheless it is still very narrow.
\par
\begin{figure}
\resizebox{0.45\textwidth}{!}{%
\includegraphics{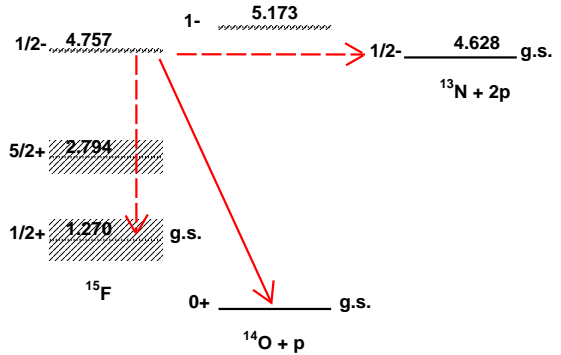}%
}
\caption{(Color online) Level scheme of $^{15}$F. The possible decay channels from the J$^{\pi}$=$1/2_1^-$ resonance are: the one proton emission (red arrow), gamma transition and two proton emission (red dashed arrow). The hatched areas correspond to the width of the resonances.}
\label{schema}
\end{figure}

There are experimental evidences in the mirror nucleus that this negative-parity state is a nearly pure $(sd)^2$ configuration coupled to the ground state of $^{13}$N \cite{Truong83,Cappuzzello12}. The emission of two protons from the narrow state is energetically possible, as seen in Fig. \ref{schema}. Since there is no intermediate state accessible to $^{14}$O, it should be a direct two-proton emission to the g.s. of $^{13}$N. However, the available energy is only Q$_{2p}$~=~129~keV, inducing a Wigner limit of $\Gamma_{^2He}$~=~4x10$^{-11}$~eV (t$_{1/2}$=16.5 $\mu$s) for the emission of a $^{2}$He cluster with $\ell$=0. Moreover, it is known that the modeling of the decay by the tunneling of a $^2$He cluster overestimates the two-proton width \cite{Grigorenko00,Grigorenko15}. Therefore, the branching ratio for the emission of two protons is expected to be extremely small.

\section{GSM Calculations}
The description of $^{15}$F requires a proper treatment of the continuum couplings. Here, we use the Gamow shell model (GSM) which provides a fully microscopic and unified description of bound and unbound nuclear states \cite{GSM,GSM1,GSM2}, and nuclear reactions \cite{Jaganathen14,Fossez15}. In the latter case, GSM is formulated in the coupled channel representation (GSM-CC).
\par
In our studies, the translationally invariant GSM Hamiltonian consists of (i) the Woods-Saxon potential with the spin-orbit term which describes the field of the $^{12}$C core acting on valence nucleons in $^{13}$N, $^{14}$O, and $^{15}$F, (ii) the Furutani-Horiuchi-Tamagaki (FHT) finite-range two-body interaction \cite{Furutani78} between valence nucleons, and (iii) the recoil term (for details see Ref. \cite{Fossez15}). Parameters of the Hamiltonian are adjusted to reproduce the binding energies of low-lying states, and the one- and two-proton separation energies in $^{15}$F. Due to the absence of the three-body interaction, parameters of the GSM Hamiltonian had to be slightly readjusted in $^{14}$O because no set of parameters of this effective two-body interaction can reproduce simultaneously spectra and binding energies of $^{13}$N, $^{14}$O, $^{15}$F.
\par
 GSM-CC calculations are performed in three resonant shells: $0p_{1/2}$, $0d_{5/2}$ and $1s_{1/2}$, and several shells in the non-resonant continuum along the discretized contours: $\mathcal{L}^+_{d_{5/2}}$ and $\mathcal{L}^+_{s_{1/2}}$ in the complex momentum $k$ plane. Each contour consists of three segments joining the points:  $k_{\rm {min}}$=0.0, $k_{\rm {peak}}=0.3-i0.1$ fm$^{-1}$,
 $k_{\rm {middle}}$=0.6 fm$^{-1}$ and $k_{\rm {max}}$=2.0 fm$^{-1}$ for $\mathcal{L}^+_{d_{5/2}}$, and $k_{\rm {min}}$=0.0, $k_{\rm {peak}}=0.25-i0.1$ fm$^{-1}$, $k_{\rm {middle}}$=0.5 fm$^{-1}$ and $k_{\rm {max}}$=2.0 fm$^{-1}$ for $\mathcal{L}^+_{s_{1/2}}$. Each segment is discretized with 10 points. The states along each contour are generated by the same WS potential. The $p_{1/2}$ continuum is approximated by 5 lowest harmonic oscillator (HO) wave functions.
Similarly, the $p_{3/2}$ and $d_{3/2}$ continua are approximated by 5 and 6 HO states, respectively. To reduce the size of the GSM matrix, the basis of Slater determinants is truncated by limiting to 2 the number of nucleons in the non-resonant continuum states.
 \par
Antisymmetric eigenstates of the GSM-CC have been expanded in the basis of channel states which are built by coupling the GSM wave functions for ground state $0^+_1$ and excited states $1^-_1$ , $0^+_2$ , $3^-_1$, $2^+_1$, $0_1^-$, $2_2^+$, $2_1^-$ of $^{14}$O with the proton wave functions in partial waves: $s_{1/2}$, $p_{1/2}$, $p_{3/2}$, $d_{3/2}$, and $d_{5/2}$. The WS potential is fitted to reproduce the level scheme of $^{13}$N.

The two-body part of the FHT interaction from which the microscopic channel-channel coupling potentials are calculated, has been rescaled by the multiplicative factors $1.07$, 0.96 and 0.95 for ${1/2}_1^+$, ${5/2}_1^+$ and ${1/2}_1^-$ states of $^{15}$F, respectively, to compensate for neglected channels built from higher lying resonances and non-resonant continuum states of $^{14}$O.
\par
The calculated binding energies in $^{15}$F are: $(E,\Gamma)^{(\rm GSM-CC)}=(-8.29~{\rm MeV},0.437~{\rm MeV})$, $(-6.66~{\rm MeV},0.211~{\rm MeV})$, and $(-4.48~{\rm MeV},0.031 {\rm~MeV})$ for $1/2_1^+, 5/2_1^+$ and $1/2_1^-$ states, respectively. All energies are given relative to the energy of $^{12}$C. In the same scale, experimental values are: $(E,\Gamma)^{(\rm exp)}=(-8.21~{\rm MeV},0.376~{\rm MeV})$, $(-6.57~{\rm MeV},0.305~{\rm MeV})$, and $(-4.48~{\rm MeV},0.036~{\rm MeV})$. We have checked consistency between the eigenvalues calculated either in the Slater determinant representation (GSM), or in the coupled channel representation (GSM-CC) of the many-body wave functions of $^{15}$F. Calculated one- and two-proton separation energies in $^{15}$F reproduce the experimental separation energies.

\par
The narrow resonance $1/2_1^-$ can decay either by 1p- or 2p-emission. The GSM wave function for this state:
\begin{eqnarray}
 {< \Psi | 0p_{1/2}[1] s_{1/2}[2]>}^2 &=& 0.97 \\ \nonumber
{< \Psi | 0p_{1/2}[1] 0d_{5/2}[2]>}^2 &=& 0.02 \nonumber
\end{eqnarray}
is an almost pure wave function of 2 protons in $s_{1/2}$ resonant and non-resonant shells.

\par
Even if the spectroscopic factors are model dependent, their values within a given theoretical framework provide an useful insight into the structure of the wave functions. For $1/2_1^+$ state, the largest one-proton spectroscopic factor  is to the ground state of $^{14}$O ($S_{\rm SF}^{(1/2^+)}=0.945$). Similarly for $5/2_1^+$ state, the largest spectroscopic factor ($S_{\rm SF}^{(5/2^+)}=0.93$) is also to the ground state of $^{14}$O. On the contrary, the largest spectroscopic factors for the $1/2_1^-$ state are to the first and second excited states $1_1^-$ and $0_2^+$ of $^{14}$O, whereas the spectroscopic factor to the ground state is only $S_{\rm SF}^{(1/2^-)}=0.0035$.

\par
The GSM-CC excitation function for the reaction $^{14}$O(p,p)$^{14}$O at 180$^{\circ}$ in the c.m. is shown in Fig. \ref{14Ocm} (green dotted line). The overall agreement with the experimental results and with of R-Matrix fit is good. The calculated cross section above the $5/2_1^+$ resonances is lower than the data. This could be explained by the absence of the unknown higher-lying resonances taken into account in the R-Matrix fit (see text).

\section{Discussion}

It is tempting to compare a pattern of the resonances in $N=Z$ nucleus ($^8$Be) and in $N<Z$ nucleus ($^{15}$F). Obviously, the sequence of decay channels in $^8$Be and $^{15}$F is quite different. Nevertheless, in both nuclei one finds resonances in the vicinity of the second decay threshold which carry an imprint of this decay channel. $S_{\rm 1p}\simeq S_{\rm 1n}$ in $^8$Be and therefore one finds close lying doublets of the strongly mixed states. This mixing can be understand in an avoided crossing scenario which leads to significant width redistribution. In $^{15}$F, $S_{\rm 1p}\ll S_{\rm 1n}$ and consequently the partner state in the doublet is at high energies and does not mix with the near-threshold $1/2_1^-$ state.
\par
Electromagnetic transitions between resonances have been rarely observed, e.g. in the unbound nucleus $^8$Be \cite{Datar05,Datar13}, and in $^{56}$Cu \cite{Orrigo14}. The second excited state of $^{15}$F being a long-lived resonance, a $\gamma-$transition from this resonance to the g.s. resonance is conceivable. $E1$ transitions occurring between $2s_{1/2}\rightarrow1p_{1/2}$ single-particle states are expected to be extremely fast. In $^{11}$Be, the $1/2_1^- \rightarrow 1/2_1^+$  $\gamma$-transition is the fastest known dipole transition between bound states. This remarkable property is due to the neutron halo structure of $^{11}$Be \cite{Millener83}. The $\gamma$-width is larger since the electric transition is proportional to the radial integral $\int u_f(r)~r~u_i(r)dr$ where $u_{f,i}(r)$ are final and initial radial wave functions of the nucleon, the ground state of $^{11}$Be having a very extended radial wave function. This is also the case of $^{15}$F. Taking this property into account and the neutron/proton effective charge difference, we predict $\Gamma_{\gamma}\approx$~50~eV.
It would be interesting to measure these $\gamma$-rays in order to elucidate the structure of this unique $1/2_1^-$ state and its 1$s_{1/2}$ content. These $\gamma$-rays will be in coincidence with protons emitted from the ground state of $^{15}$F.
Other narrow resonances in $^{15}$F which were predicted in $^{15}$F at higher excitation energies \cite{Canton06,Fortune07} remain to be observed. The existence of relatively narrow resonances at high excitation energies may be actually more frequent than thought opening a possibility for the particle and gamma resonance spectroscopy in nuclei far beyond the drip lines.

\section{Conclusion}

In summary, the resonant elastic scattering technique was employed with a post-accelerated radioactive beam of $^{14}$O in order to study the structure of the $^{15}$F low-lying states. Thanks to the high beam quality, an excellent energy resolution was obtained. The second excited state of $^{15}$F with spin J$^{\pi}=1/2_1^-$ was observed as a narrow resonance located above the Coulomb barrier.
\par
The structure of the observed resonances has been analyzed using GSM. In both models which include the coupling to the continuum of decay channels and scattering states, the narrow resonance ($1/2_1^-$) is dominated by a diproton configuration. The non-resonant continuum $s_{1/2}$ plays an essential role in the structure of this state. The diproton nature of $1/2_1^-$ state implies that the 1p decay width is suppressed as compared to widths of low-lying levels $1/2_1^+$ and $5/2_1^+$.

\section{Acknowledgements}
\par
We thank the GANIL accelerator staff for delivering the radioactive beams, and A. Navin, B. Bastin and R.J. de Boer for their help and interesting discussions. This work has been supported in part by the European Community FP6 - Structuring the ERA - Integrated Infrastructure Initiative- contract EURONS n° RII3-CT-2004-506065, by the COPIN and COPIGAL French-Polish scientific exchange programs, by the France-Romanian IN2P3-IFIN-HH collaboration No. 03-33, by the France-Czech LEA NuAG collaboration, and by the Helmholtz Association (HGF) through the Nuclear Astrophysics Virtual Institute (NAVI). Some of us (P.U., I.C., K.S.) acknowledge the support by the French-Serbian collaboration agreement (No. 20505) and MESTD of Serbia (Project No.171018).

\end{document}